\documentclass[12pt,prl,hyperref,groupedaddress,nofootinbib]{revtex4}
\pdfoutput=1
\usepackage{color}
\usepackage{amsfonts}
\usepackage{amsmath}
\usepackage{dsfont}
\usepackage{amssymb}
\usepackage{graphicx}
\usepackage{subfigure}
\usepackage{psfrag}
\providecommand{\U}[1]{\protect\rule{.1in}{.1in}}
\providecommand{\U}[1]{\protect\rule{.1in}{.1in}}

\begin{document}

\def\blue#1{{\color{blue}{#1}}}

\title{Conformal bootstrap for percolation and polymers}

\author{Andr\'e LeClair}
\affiliation{
Department of Physics, Cornell University, Ithaca, NY}

\author{Joshua Squires}
\affiliation{
Department of Physics, Cornell University, Ithaca, NY}

\begin{abstract}
\qquad

The conformal bootstrap is applied to percolation and dilute self-avoiding polymers in arbitrary dimension $D$.    In both cases we propose a spectrum of operators motivated by Virasoro symmetry  in $D=2$ which is  devoid of a stress energy tensor as an approximate means of enforcing $c=0$. Percolation is treated in $2\leq D \leq 6$ dimensions, and the self-avoiding walk in $2 \leq D \leq 4$. 

\end{abstract}

\maketitle

%
%
%
\def\oti{{\otimes}}
\def\lb{ \left[ }
\def\rb{ \right]  }
\def\tilde{\widetilde}
\def\bar{\overline}
\def\hat{\widehat}
\def\*{\star}
\def\[{\left[}
\def\]{\right]}
\def\({\left(}
\def\BL{\Bigr(}
\def\){\right)}
\def\BR{\Bigr)}
\def\BBL{\lb}
\def\BBR{\rb}
\def\zb{{\bar{z} }}
\def\zbar{{\bar{z} }}
\def\frac#1#2{{#1 \over #2}}
\def\inv#1{{1 \over #1}}
\def\half{{1 \over 2}}
\def\d{\partial}
\def\der#1{{\partial \over \partial #1}}
\def\dd#1#2{{\partial #1 \over \partial #2}}
\def\vev#1{\langle #1 \rangle}
\def\bra#1{{\langle #1 |  }}
\def\ket#1{ | #1 \rangle}
\def\rvac{\hbox{$\vert 0\rangle$}}
\def\lvac{\hbox{$\langle 0 \vert $}}
\def\2pi{\hbox{$2\pi i$}}
\def\e#1{{\rm e}^{^{\textstyle #1}}}
\def\grad#1{\,\nabla\!_{{#1}}\,}
\def\dsl{\raise.15ex\hbox{/}\kern-.57em\partial}
\def\Dsl{\,\raise.15ex\hbox{/}\mkern-.13.5mu D}

\def\th{\theta}		
\def\Th{\Theta}
\def\ga{\gamma}		
\def\Ga{\Gamma}
\def\be{\beta}
\def\al{\alpha}
\def\ep{\epsilon}
\def\vep{\varepsilon}
\def\la{\lambda}	
\def\La{\Lambda}
\def\de{\delta}		
\def\De{\Delta}
\def\om{\omega}		
\def\Om{\Omega}
\def\sig{\sigma}	
\def\Sig{\Sigma}
\def\vphi{\varphi}

\def\CA{{\cal A}}	\def\CB{{\cal B}}	\def\CC{{\cal C}}
\def\CD{{\cal D}}	\def\CE{{\cal E}}	\def\CF{{\cal F}}
\def\CG{{\cal G}}	\def\CH{{\cal H}}	\def\CI{{\cal J}}
\def\CJ{{\cal J}}	\def\CK{{\cal K}}	\def\CL{{\cal L}}
\def\CM{{\cal M}}	\def\CN{{\cal N}}	\def\CO{{\cal O}}
\def\CP{{\cal P}}	\def\CQ{{\cal Q}}	\def\CR{{\cal R}}
\def\CS{{\cal S}}	\def\CT{{\cal T}}	\def\CU{{\cal U}}
\def\CV{{\cal V}}	\def\CW{{\cal W}}	\def\CX{{\cal X}}
\def\CY{{\cal Y}}	\def\CZ{{\cal Z}}


\def\barray{\begin{eqnarray}}
\def\earray{\end{eqnarray}}
\def\beq{\begin{equation}}
\def\eeq{\end{equation}}

\def\Tr{\rm Tr} 
\def\xvec{{\bf x}}
\def\kvec{{\bf k}}
\def\kvecp{{\bf k'}}
\def\omk{\om{\kvec}} 
\def\dk#1{\frac{d\kvec_{#1}}{(2\pi)^d}}
\def\2pid{(2\pi)^d}
\def\ket#1{|#1 \rangle}
\def\bra#1{\langle #1 |}
\def\vol{V}
\def\adag{a^\dagger}
\def\rme{{\rm e}}
\def\Im{{\rm Im}}
\def\pvec{{\bf p}}
\def\fermiS{\CS_F}
\def\cdag{c^\dagger}
\def\adag{a^\dagger}
\def\bdag{b^\dagger}
\def\vvec{{\bf v}}
\def\muhat{{\hat{\mu}}}
\def\vac{|0\rangle}
\def\pcut{{\Lambda_c}}
\def\chidot{\dot{\chi}}
\def\gradvec{\vec{\nabla}}
\def\psitilde{\tilde{\Psi}}
\def\psibar{\bar{\psi}}
\def\psidag{\psi^\dagger} 
\def\m{m_*}
\def\up{\uparrow}
\def\down{\downarrow}
\def\Qo{Q^{0}}
\def\vbar{\bar{v}}
\def\ubar{\bar{u}}
\def\smallhalf{{\textstyle \inv{2}}}
\def\smallsqrt{{\textstyle \inv{\sqrt{2}}}}
\def\rvec{{\bf r}}
\def\avec{{\bf a}}
\def\pivec{{\vec{\pi}}}
\def\svec{\vec{s}} 
\def\phivec{\vec{\phi}}
\def\daggerc{{\dagger_c}}
\def\Gfour{G^{(4)}}
\def\dim#1{\lbrack\!\lbrack #1 \rbrack\! \rbrack }
\def\qhat{{\hat{q}}}
\def\ghat{{\hat{g}}}
\def\nvec{{\vec{n}}}
\def\bull{$\bullet$}
\def\ghato{{\hat{g}_0}}
\def\r{r}
\def\deltaq{\delta_q}
\def\gcharge{g_q}
\def\gspin{g_s}
\def\deltas{\delta_s}
\def\gQC{g_{AF}} 
\def\ghatqc{\ghat_{AF}}
\def\xqc{x_{AF}}
\def\mhat{\hat{m}}
\def\xup{x_2}
\def\xdown{x_1}
\def\sigmavec{\vec{\sigma}}
\def\xopt{x_{\rm opt}}
\def\Lambdac{{\Lambda_c}}
\def\angstrom{{{\scriptstyle \circ} \atop A}     }
\def\AA{\leavevmode\setbox0=\hbox{h}\dimen0=\ht0 \advance\dimen0 by-1ex\rlap{
\raise.67\dimen0\hbox{\char'27}}A}
\def\ratio{\gamma}
\def\Phivec{{\vec{\Phi}}}
\def\singlet{\chi^- \chi^+} 
\def\mhat{{\hat{m}}}
\def\blue#1{{\color{blue}{#1}}}
\def\red#1{{\color{red}{#1}}}
\def\Im{{\rm Im}}
\def\Re{{\rm Re}}
\def\xstar{x_*}
\def\sech{{\rm sech}}
\def\Li{{\rm Li}}
\def\dim#1{{\rm dim}[#1]}
\def\ep{\epsilon}
\def\free{\CF}
\def\Fhat{\digamma}
\def\ftilde{\tilde{f}}
\def\muphys{\mu_{\rm phys}}
\def\xiprime{\tilde{\xi}}
\def\CI{\mathcal{I}}
\def\ko{k_0}
\def\Lambdastar{\Lambda_*}   
\def\gtilde{\tilde{g}} 
\def\ntilde{\tilde{n}}
\def\mutilde{\tilde{\mu}}

\section{I. Introduction}

The conformal bootstrap is the idea that a conformally invariant quantum field theory is completely characterized by its spectrum of anomalous dimensions and operator product expansion coefficients \cite{Polyakov}.     In $D=2$  dimensions,    implementation of the bootstrap is  hardly necessary since  the conformal symmetry  becomes the infinite dimensional Virasoro  symmetry,  which leads to powerful methods such as Coulomb gas techniques,  current algebra and their cosets,  etc.  \cite{CFTbook}.   
Remarkably,   recently it has been demonstrated that the conformal bootstrap can provide  accurate results in higher dimensions \cite{Rattazzi}.   In particular for the $D=3$ Ising model,  
the best results on anomalous dimensions is currently based on the bootstrap \cite{Kos2}.     For reviews see \cite{SimmonsDuffin,Rychkov}.  

In this paper we explore the power,  or possible limitations,  of the bootstrap for two conformal theories that are as important as the Ising model,   namely percolation and polymers.    
The latter is commonly referred to as the self-avoiding walk (SAW).     These theories present several interesting challenges  in the context of the conformal bootstrap.  First of all,  they are not unitary.   Furthermore,  they are very closely related in that they share some anomalous dimensions,  and in $D=2$ they have the same  Virasoro central charge $c=0$.  It should be mentioned that some important problems in Anderson localization,  such as the critical point in quantum Hall transitions for non-interacting fermions,    are  also expected 
to be described by $D=2$,  $c=0$ conformal field theories,  many of whose  description remains unknown.     In contrast,   the Ising model is essentially a unique theory:  in $D=2$ it is the unique unitary theory with central charge $c=1/2$,  which makes it easier to locate.      In light of these comments,  the main goal of this article  is to explore whether the conformal bootstrap can distinguish between percolation and the SAW in any dimension $D$.   As we will  argue,  the answer is affirmative.    Our goal is not to provide highly accurate numerical results for conformal exponents,  but rather to simply  argue that the bootstrap is powerful enough to locate these two theories,   however in a subtle way.    We provide numerical estimates of exponents based on our proposal which are reasonably good,  however not as accurate as those obtained by other methods such as $\epsilon$-expansion or Monte-Carlo,  although our results can probably be improved with more extensive numerical studies.

In order to describe the problem,  and establish notation,  let us consider the $D=2$ case where exact results are known.   The unitary minimal models have
central charge 
\beq 
\label{c} 
c = 1 - \frac{6}{p(p+1)} \geq 1/2
\eeq
     They contain primary fields $\Phi_{r,s}$,  with $1\leq s \leq p$, $1\leq r \leq p-1$ with scaling dimension 
\beq
\label{Deltars}
\Delta_{r,s}=  2 h_{r,s}= \frac{ \( (p+1) r - p s\)^2 -1}{2 p (p+1)}
\eeq
For concreteness consider the Ising model at $p=3$ with $c=1/2$.      The model can be perturbed away from it's critical point by either changing the temperature away from the critical temperature $T_c$ and or turning on a magnetic field.     One is thus led to consider the action 
\beq
\label{action} 
S = S_{\rm cft}  + \int d^D x \Bigl(  g_t  \, \epsilon (x)  + g_m \,  \sigma (x) \Bigr)
\eeq
where  $S_{\rm cft}$ is formally the action for the conformal field theory,  $\epsilon (x)$ is the energy operator,  $\sigma (x)$ is the spin field,  and the $g$'s are couplings, 
where $g_t = T- T_c$.    It is well-known that the energy operator corresponds to $(r,s) = (2,1)$ with
$\Delta_\epsilon = 1$.     The spin field corresponds to $(r,s) = (1,2)$ with $\Delta_\sigma = 1/8$.  
They satisfy the fusion rule
\beq
\label{fusion} 
[\sigma] \times [\sigma] = [1]+ [\epsilon]
\eeq
An important exponent is the correlation length exponent $\nu$.    The dimension of the coupling $g_t$ is $D-\Delta_\epsilon$,   therefore 
$\xi = (g_t)^{-1/(D-\Delta_\epsilon)}$ has units of length and diverges as $g_t \to 0$,   thus 
\beq
\label{nudef}
\nu =  \inv{D - \Delta_\epsilon}
\eeq
For the Ising model,  $\nu = 1$.    

Consider now lowering $p$ by $1$ to $p=2$ where one encounters the first non-unitary theories at $c=0$.   The space of $c=0$ theories is vast;  in fact it is infinite.  
For instance current algebras based on the super Lie algebras $gl(n|n)$ or $osp(2n|2n)$ all have $c=0$ and have important applications to disordered systems.   
In order to limit our attention to percolation and the SAW,   we can view them as continuous limits of other models that pass through the Ising model.  
The SAW is known to correspond to the $O(N)$ model as $N\to 0$,  where Ising is $N=1$.    On the other hand percolation is the $q\to 1$ limit of the q-state Potts model, 
where the Ising model is $q=2$.     Due to these limits,  both these theories have an energy operator and spin field.   These $D=2$ theories have been extensively studied, for instance in 
\cite{CardyPerc,DotsenkoFateev,Saleur,Delfino,Delfino2,Dotsenko}.    It is known that for both theories,  the spin field corresponds 
to $(r,s) = (3/2, 3/2)$ with $\Delta_\sigma = 5/48$.    Thus,  percolation and the SAW differ in the energy sector.   For the SAW,   the energy operator 
corresponds to $(r,s) = (1,3)$ with $\Delta_\epsilon = 2/3$,  which gives $\nu = 3/4$.   On the other hand,  for percolation it is $(r,s)= (2,1)$ with dimension $\Delta_\epsilon = 5/4$ which leads to $\nu = 4/3$.

The above discussion leads to some interesting questions.    First of all,  both percolation and the SAW have the same fusion rule \eqref{fusion} and same central charge $c=0$.   
Can the conformal bootstrap deal with these important non-unitary theories?  Can it distinguish between percolation and the SAW?    Finally,   how well does it work in dimensions $D \geq  2$?  
Based on the above discussion,  we expect them to differ in the energy sector,  namely which descendants are included in $[1] + [\epsilon]$.    
In the sequel, we will propose some selection rules that appear to answer these questions.     

It should be mentioned that a detailed study of the difference between percolation and SAW in two dimensions was carried out by Gurarie and Ludwig \cite{Gurarie}.
It is known that if  $ \phi (z)$  is the holomorphic part of a primary field of weight $\Delta = 2h$,  then  one has the operator product expansion (OPE) 
\beq
\label{catas}
\phi (z) \phi (0) = \inv{z^{2h}} \( 1 + \frac{2h}{c} z^2 T(0) \)  \ldots
\eeq
where $T(z)$ is the stress energy tensor.  
Note the ``catastrophe" for $c=0$ \cite{Cardy}.    It was proposed that this can be resolved by the existence of another field $t(z)$ of weight $2$ which is the logarithmic partner to $T(z)$.   
For our purposes,  these facts will in part motivate our selection rules for the bootstrap,  in particular for the descendants of the identity,  like $T(z)$.    However we will not incorporate potential constraints from the structure of logarithmic conformal field theories in the bootstrap.

There is another important and subtle point in trying to bootstrap these theories. For both theories in 2D, the identity decouples exactly when $q=1$ or $N=0$ \cite{Delfino3,Dotsenko,Cardy}, altering the fusion rule  \eqref{fusion} to
\beq
\label{newfuse}
[\sigma] \times [\sigma] = [\epsilon].
\eeq	
This is not surprising,  since when $q=1$ or $N=0$,  the spin field does not formally exist,  which is consistent with the fact that the fusion rule \eqref{newfuse} implies that the two point function of spin fields  formally vanishes.  
 Taking percolation for example, this can be understood by noting that the probability $P$ that two sites are both contained in the same connected cluster is given by
\begin{equation}
P=\lim_{q\rightarrow 1}(q-1)^{-1}\langle \sigma(z_1)\sigma(z_2)\rangle
\end{equation} \cite{Delfino3,Cardy}. Since $P$ must be finite the two-point function must be proportional to $(q-1)$ and therefore go to zero at $q=1$.  Furthermore,  the vanishing of the identity channel is demonstrated by the more sophisticated calculation of Dotsenko \cite{Dotsenko},  through a careful renormalization procedure within the Coulomb gas formalism.
In particular,  Dotsenko had to introduce a small parameter $\epsilon$,   where $c\propto \epsilon$.   Only after he renormalized the 4-point function in a particular manner did the identity channel vanish as $\epsilon \to  0$.       In contrast, since we don't rigorously impose $c=0$, our proposed fusion rule for percolation necessarily includes the identity operator. The justification and consequences of this decision are explored in Appendix A.

This paper is organized as follows.   In the next section we review some standard methods of the conformal bootstrap.    The following two sections treat percolation and the SAW separately,  where we provide numerical evidence for our choice of selection rules for $2<D <6$.

\section{II. Conformal Bootstrap}
At the heart of the conformal bootstrap is the notion that constraints on the four-point functions of a CFT, namely conformal invariance, crossing symmetry, and unitarity, are sufficient to restrict, or even completely fix, the spectrum of allowed scaling dimensions of a theory. Conformal invariance constrains the four-point function of a scalar field $\sigma(x)$ in a CFT to take the form
\begin{equation}
\langle \sigma(x_1)\sigma(x_2) \sigma(x_3)\sigma(x_4)\rangle = \frac{\sum_{\Delta,l}p_{\Delta,l}G_{\Delta,l}(u,v)}{|x_{12}|^{2\Delta_{\sigma}}|x_{34}|^{2\Delta_{\sigma}}},
\label{fourpoint}
\end{equation} with $x_{ij} \equiv x_i-x_j$ and $\Delta_\sigma$ the scaling dimension of $\sigma$. The coefficients $p_{\Delta,l}$  are the square of the $\sigma(x_i)\sigma(x_j)$ OPE coefficients $\lambda_{\sigma \sigma \mathcal{O}}$, with $\mathcal{O}$ signifying a global primary operator of dimension $\Delta$ and conformal spin $l$. $G_{\Delta,l}(u,v)$ are global conformal blocks, which are functions of the conformally invariant cross ratios $u=\frac{x_{12}^2x_{34}^2}{x_{13}^2x_{24}^2}$ and $v=\frac{x_{14}^2x_{23}^2}{x_{13}^2x_{24}^2}$. Crossing symmetry is imposed by considering the transformation of \eqref{fourpoint} under $x_1 \leftrightarrow x_3$. Defining
\begin{equation}
F_{\Delta_{\sigma},\Delta,l}  \equiv v^{\Delta_\sigma}G_{\Delta,l}(u,v)-u^{\Delta_\sigma}G_{\Delta,l}(v,u)
\end{equation} crossing symmetry is respected if
\begin{equation}
\sum_{\Delta,l}p_{\Delta,l}F_{\Delta_{\sigma},\Delta,l} = 0.
\label{crossing}
\end{equation}

In unitary theories, the coefficients $p_{\Delta,l}$ are strictly positive due to reality of $\lambda_{\sigma \sigma \mathcal{O}}$. The contemporary conformal bootstrap \cite{Rattazzi},  only took shape after crucial advances in the study of conformal blocks \cite{Dolan1,Dolan2}.  It has since been refined and applied most notabably to the $O(N)$ models \cite{Gopakumar,Shimada, ElShowk1,ElShowk2,ElShowk3,Kos1,Kos2,Kos3,Kos4,Rattazzi2,Rattazzi3,Iliesiu,Alday1,Alday2}.    In this approach,  a  functional $\Lambda$ is sought such that $\Lambda(F_{\Delta_{\sigma},\Delta,l}) \geq 0$. When this condition is satisfied, it contradicts the crossing relation \eqref{crossing} since $p_{\Delta,l}>0$. Therefore regions of parameter space where such a $\Lambda$ exists cannot correspond to a physical CFT, and bounds can be placed on the possible scaling dimensions.

In the absence of unitarity, an alternate formulation of the conformal bootstrap which does not rely on the positivity of $p_{\Delta,l}$ is required. In the determinant or ``Gliozzi" conformal bootstrap method \cite{Gliozzi1,Gliozzi2}, this requirement is eliminated at the expense of generality. Rather than searching the space of all possible CFTs for bounds which are independent of a specific theory, a particular CFT must be chosen beforehand by specifying the dimensions and conformal spins of the first $N$ operators that appear in the crossing relation. This method has been applied to  the Yang-Lee edge singularity \cite{Gliozzi1,Gliozzi2,Hikami1} and polymers \cite{Hikami2}. To set up this approach, we perform the standard variable change $v = ((2-a)^2-b)/4$, $u=(a^2-b)/4$ and Taylor expand \eqref{crossing} around $a=1,b=0$, generating the homogeneous system
\begin{equation}
\sum_{\Delta,l}p_{\Delta,l} F^{(m,n)}_{\Delta_{\sigma},\Delta,l}=0 \qquad (m,n \in \mathbb{N},m \, \text{odd})
\label{constraint}
\end{equation} where
\begin{equation}
F^{(m,n)}_{\Delta_{\sigma},\Delta,l} = \partial^m_a \partial^n_b \left(v^{\Delta_\sigma}G_{\Delta,l}(u,v)-u^{\Delta_\sigma}G_{\Delta,l}(v,u)\right)|_{a,b=1,0}. 
\label{F}
\end{equation} Note the exclusion of even $m$ is owed to the two terms of \eqref{F} contributing oppositely in such cases. Truncating the sum to the first $N$ operators appearing in the OPE and taking $M\geq N$ derivatives, where each $M$ signifies a distinct $(m,n) = \partial^m _a \partial^n _b$ pair, gives a system of $\begin{pmatrix} M \\ N \end{pmatrix}$ equations which has a solution only if all minors of order $N$ vanish. Instead of searching for intersections of vanishing minors, we adopt the equivalent condition \cite{Esterlis} that the $M \times N$ matrix $\mathbf{F}$, with elements $F^{(m,n)}_{\Delta_\sigma,\Delta,l}$, must have at least one vanishing singular value. 

As more operators are kept in the truncation of \eqref{crossing}, additional derivatives must be added. For smaller matrices the set of derivatives chosen can greatly influence the bootstrapped scaling dimensions, as explored in the appendix. This appears to be an inherent ambiguity in the determinant conformal bootstrap method (hereafter referred to simply as the conformal bootstrap), and a method of objectively choosing derivatives should be decided. We find using only longitudinal derivatives $(m,0)$ to be most effective. Calculation of $F^{(m,n)}_{\Delta_\sigma,\Delta,l}$ is performed with the numerical bootstrap package JuliBootS \cite{Juliboots}, which implements a partial fraction representation of conformal blocks \cite{ElShowk2} and recursively calculates their derivatives \cite{Hogervorst}. Finally, before describing how the conformal bootstrap is applied to percolation and the SAW, we note that while in this work no attempt is made to calculate the error introduced in truncating \eqref{crossing}, a recent study \cite{Li} has taken steps to formalize an error estimation procedure.

\section{III. Percolation}

Let us first provide some arguments for the selection rules we will impose on the operator content of the conformal bootstrap.    It is not difficult to show that there is a null state at level $2$ for a primary field with conformal weights $h$, $\overline{h}$ if the following equation is satisfied
\beq
\label{null}
c = \frac{2h(5-8h)}{(2h+1)}
\eeq
(See  for instance \cite{Ginsparg}.)
For $c=0$,  this null state occurs at $h=5/8$ and $h=0$.   Since $h=5/8$ corresponds to the energy operator,  this suggests we discard its level 2 descendant, $[\Delta_\epsilon+2, 2 ]$. Here we introduce the notation $[\Delta,l]$ to represent an operator with dimension $\Delta = h+\overline{h}$ and conformal spin $l=h-\overline{h}$.
The $c=0$ catastrophe discussed in relation to equation \eqref{catas} also suggests we discard $[D,2]$   and its descendants,  based on the null state at $h=0$.
One can also interpret this as effectively setting $T=0$ in \eqref{catas} to avoid the $c=0$ catastrophe.   This motivates  a fusion rule consisting of the identity operator and Virasoro descendants of $\epsilon$:
 \begin{equation}
 [\Delta_\sigma,0] \times [\Delta_\sigma,0] =  [0,0] + [\Delta_\epsilon,0] + [\Delta_\epsilon+4,4] + [\Delta_\epsilon+6,6] + [\Delta_\epsilon+8,8]+ \dots
 \label{perc_spectrum}
 \end{equation} 
 
 Curiously, when constructing $\mathbf{F}$ with the above operators we observe a noticeable increase in the accuracy of the bootstrapped 2D percolation scaling dimensions if the  $(m,n)=(1,0)$ constraint is avoided.  For consistency we also omit the $(1,0)$ derivative constraint in higher dimensions, as well as in our treatment of the self-avoiding walk.  In all dimensions considered for percolation, the $M$ rows of $\mathbf{F}$ are labeled by the $M$ lowest order longitudinal derivatives with $m\geq 3$, and the $N$ columns are labeled by the first $N$ operators present in the trial spectrum \eqref{perc_spectrum}. A discussion of the decision to use only longitudinal derivatives with $m\geq 3$  is provided in Appendix B.
 
\begin{figure}
\begin{center}
\includegraphics[width=0.65\textwidth]{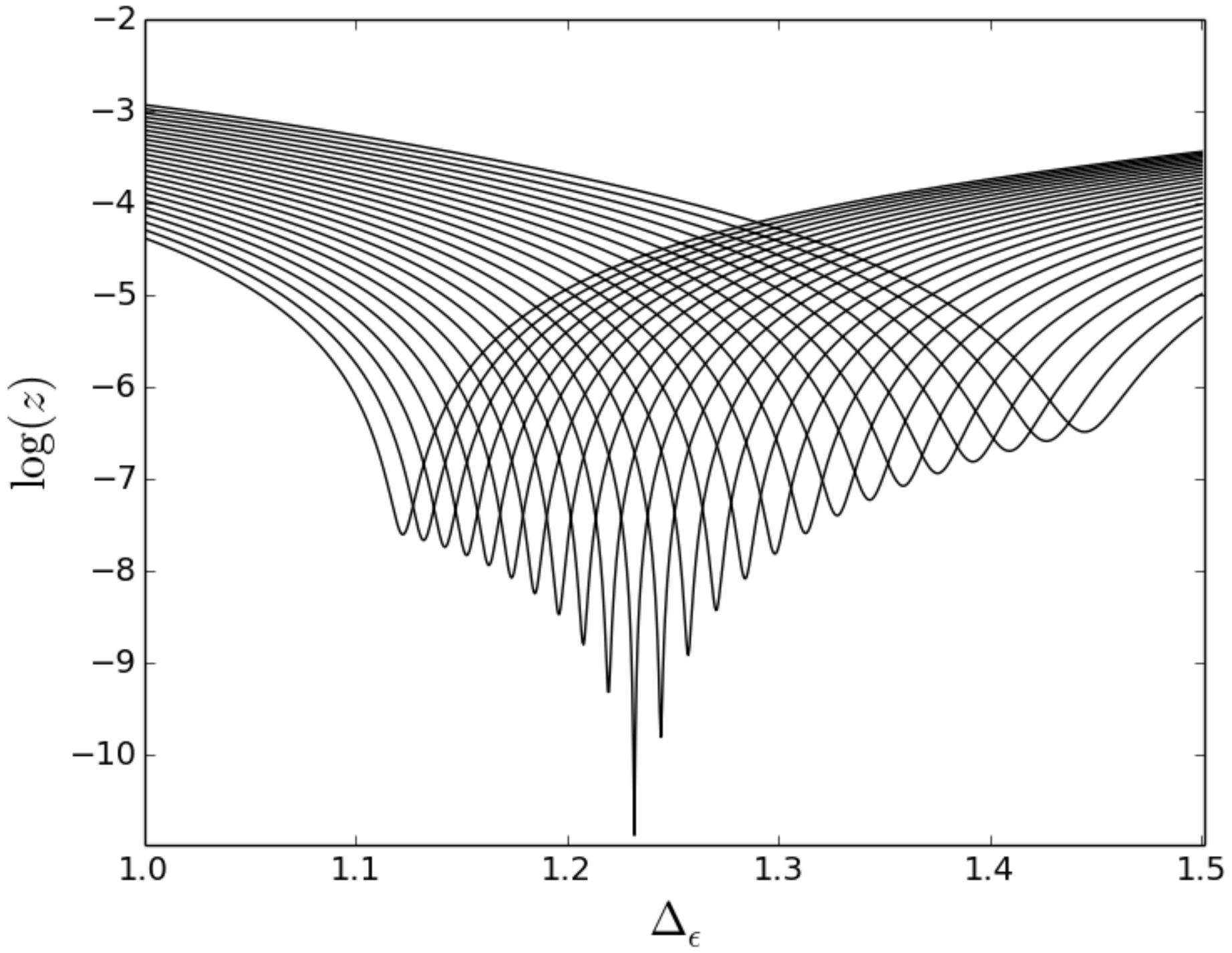}
\caption{{\bf 2D Percolation.}   Logarithm  of the smallest singular value $z$ of $\mathbf{F}$ with $N=5, M=6$ as a function of $\Delta_\sigma$ and $\Delta_\epsilon$. Each curve corresponds to a distinct value of $\Delta_\sigma$, linearly spaced from $\Delta_\sigma = 4/48$ (left-most dip) to $\Delta_\sigma =6/48$ (right-most dip). The minimal $\log(z)$ occurs at $\Delta_\sigma = 0.101, \Delta_\epsilon=1.235$. }
\label{2DP}
\end{center}
\end{figure}

\begin{figure}
\begin{center}
\includegraphics[width=0.65\textwidth]{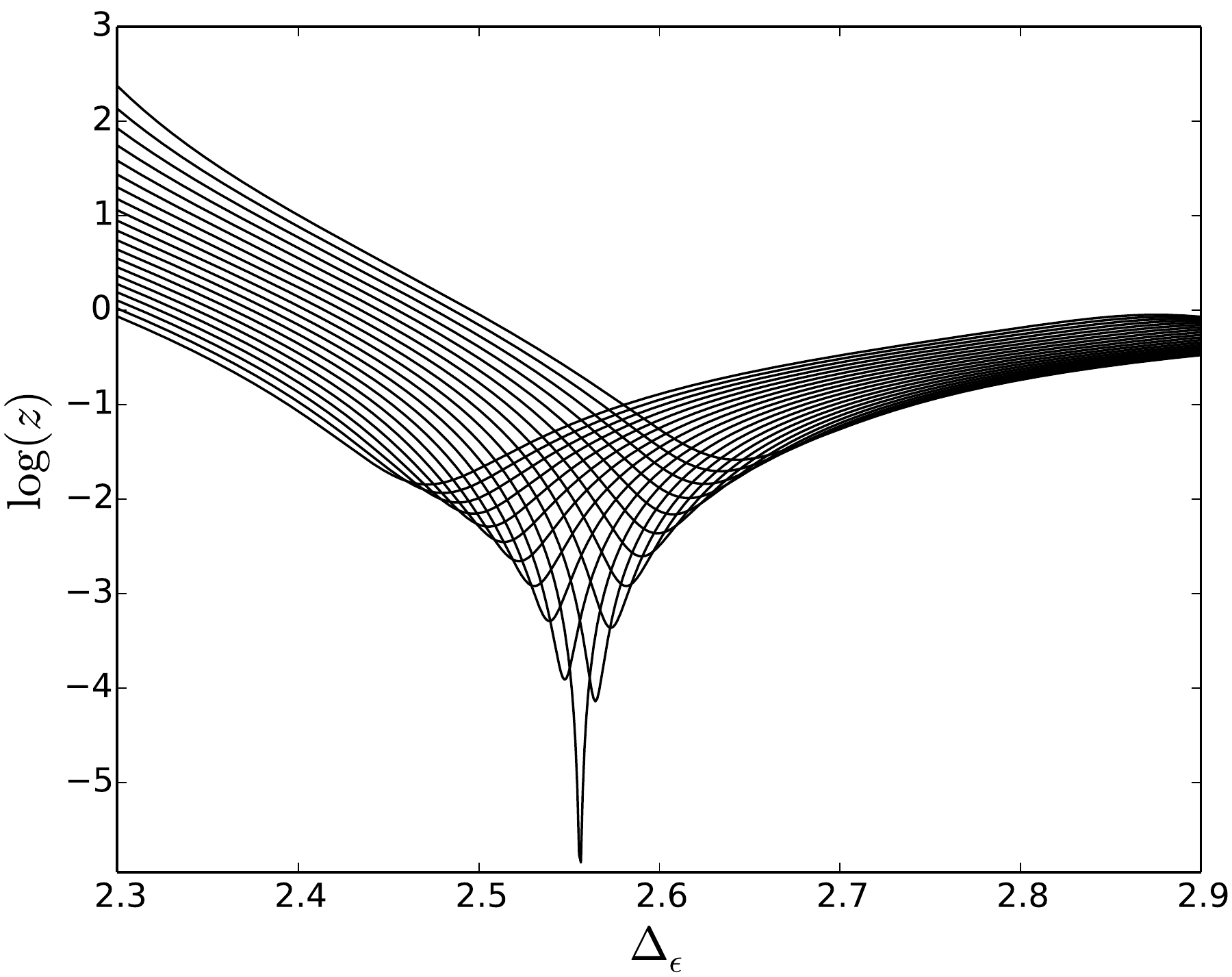}
\caption{{\bf 4D Percolation.}  Logarithm  of the smallest singular value $z$ of the matrix  $\mathbf{F}/F^{(3,0)}_{\Delta_\sigma,0,0}$ with $N=5, M=6$ as a function of $\Delta_\sigma$ and $\Delta_\epsilon$. Each curve corresponds to a distinct value of $\Delta_\sigma$, linearly spaced from $\Delta_\sigma = 0.98$ (left-most dip) to $\Delta_\sigma =1.02$ (right-most dip). The minimal $\log(z)$ occurs at $\Delta_\sigma = 0.997, \Delta_\epsilon=2.557$. }
\label{4DP}
\end{center}
\end{figure}

 Bootstrapping in $D=2$ dimensions with the above operators for fixed $\Delta_\sigma=5/48$ gives a vanishing singular value at $\Delta_\epsilon=1.255$, in agreement with the exact $\Delta_\epsilon=5/4$. Varying the spin field scaling dimension and minimizing $z$, the smallest singular value of $\mathbf{F}$, as a function of both $\Delta_\epsilon$ and $\Delta_\sigma$ finds $\Delta_\sigma = 0.101, \Delta_\epsilon=1.235$, as shown in Figure \ref{2DP}. In $D=4$ the presence of the free field theory with scaling dimensions  $\Delta_\sigma=1$ and $\Delta_\epsilon=2$ makes it difficult to minimize in $\Delta_\sigma$, and the omission of the $(1,0)$ derivative constraint only compounds the problem. All higher order derivatives of the convolved vacuum conformal block $F_{\Delta_\sigma,0,0}$ quickly tend to zero as the free field $\Delta_\sigma$ is reached, since $F^{(1,0)}_{\Delta_\sigma,0,0} $ becomes linear as $\Delta_\sigma \rightarrow 1$. Thus with our approach a trivial vanishing singular value near $\Delta_\sigma=1$ is unavoidable in four dimensions. Nevertheless, minimizing the smallest singular value of $\mathbf{F}$ gives $\Delta_\sigma=0.997, \Delta_\epsilon=2.557$. This solution is depicted in Figure \ref{4DP}, where we actually work with the scaled matrix $\mathbf{F}/F^{(3,0)}_{\Delta_\sigma,0,0}$. This is purely for visual convenience; it smooths the precipitous dip in $z$ near $\Delta_\sigma=1$ but has no bearing on the bootstrapped scaling dimensions. Our bootstrapped $\Delta_\epsilon$ corresponds to a correlation length critical exponent $\nu=0.693$ which compares favorably with $\nu=0.6920$, obtained by four-loop calculation \cite{Gracey}.
 
\begin{figure}
\begin{center}
\includegraphics[width=0.65\textwidth]{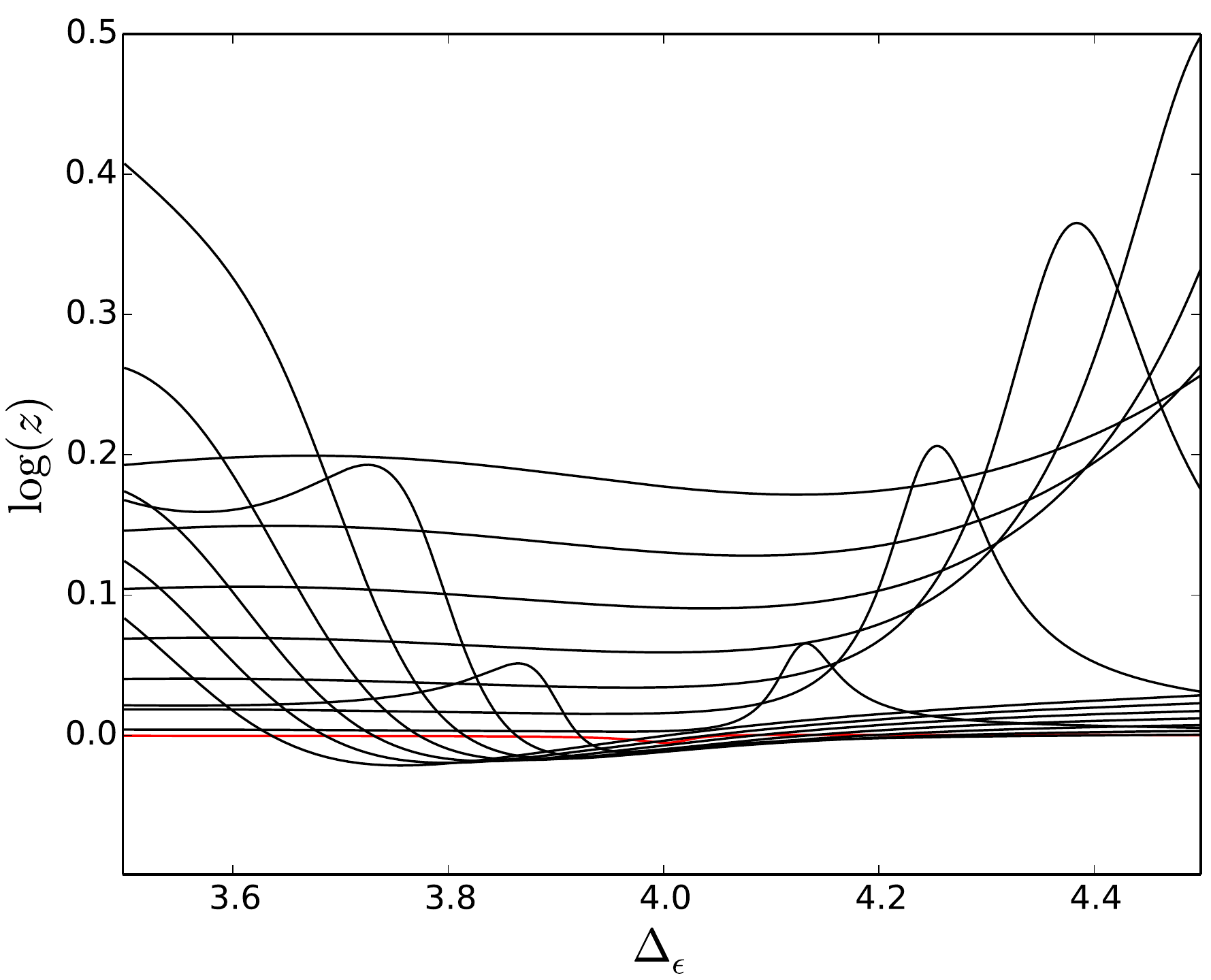}
\caption{{\bf 6D Percolation.}   Logarithm  of the smallest singular value $z$ of the matrix  $\mathbf{F}/F^{(3,0)}_{\Delta_\sigma,0,0}$ with $N=6, M=8$ as a function of $\Delta_\sigma$ and $\Delta_\epsilon$. Each curve corresponds to a distinct value of $\Delta_\sigma$, linearly spaced from $\Delta_\sigma = 1.8$ (left) to $\Delta_\sigma =2.2$ (right). Near $\Delta_\epsilon=4$ the curves flatten. The red curve corresponding to $\Delta_\sigma=2.002$ has the smallest peak and a minimum at $\Delta_\epsilon=4.003$.  As in $4D$, using $\mathbf{F}/F^{(3,0)}_{\Delta_\sigma,0,0}$ rather than $\mathbf{F}$ has no bearing on the determination of the spin and energy operator scaling dimensions.}
\label{6DP}
\end{center}
\end{figure}

\begin{figure}
\begin{center}
\includegraphics[width=0.65\textwidth]{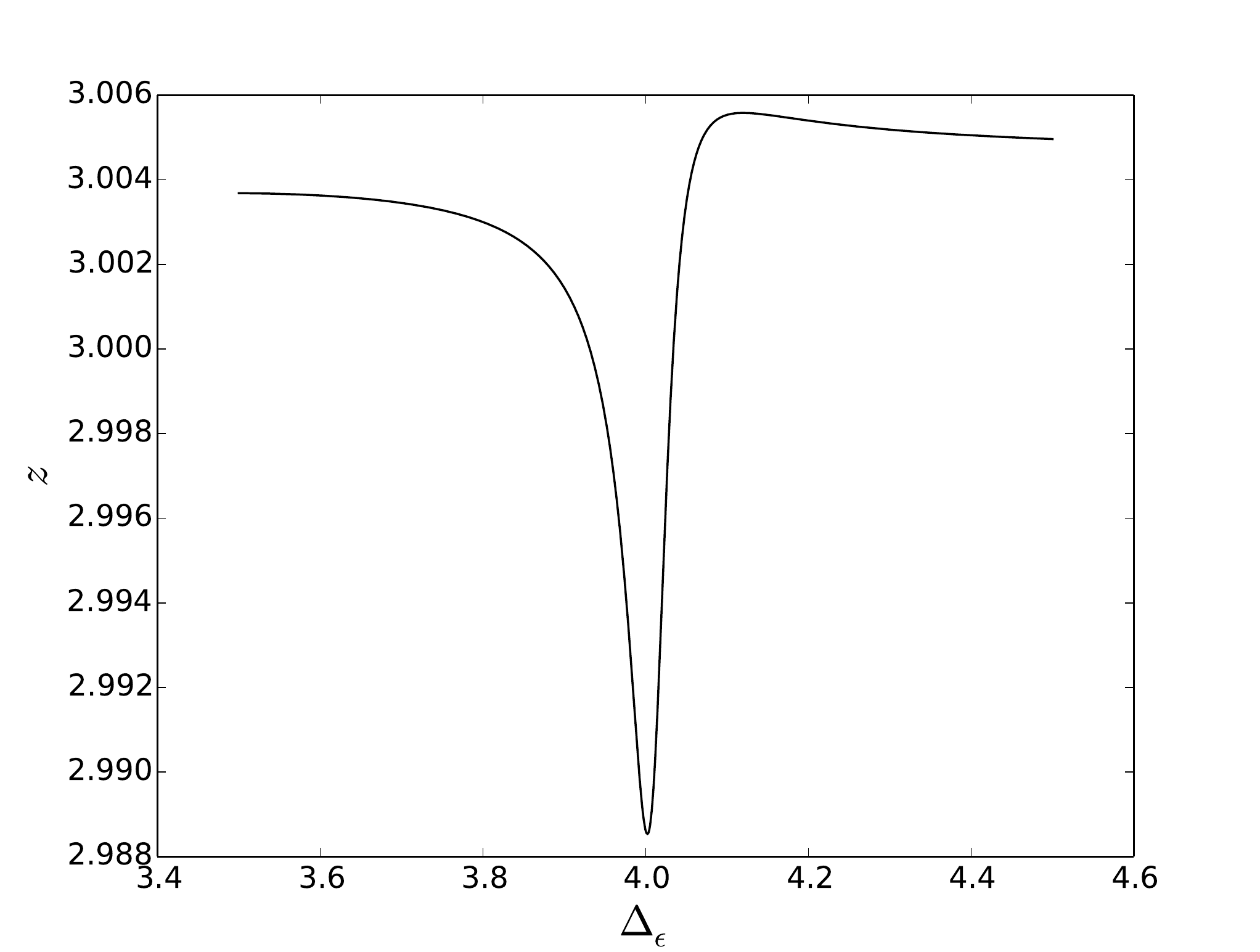}
\caption{{\bf 6D Percolation.}  Smallest singular value $z$ of $\mathbf{F}$ at fixed $\Delta_\sigma=2.002$ (red curve from Figure \ref{6DP}) achieves its minimal value at $\Delta_\epsilon=4.003$.}
\label{6DP_2}
\end{center}
\end{figure}

Applying the bootstrap to percolation's upper critical dimension $D=6$ with the same OPE truncation as in two and four dimensions is unsuccessful. No vanishing singular values of $\mathbf{F}$ are found when $M>N$, which for our minimal set of operators appears to be necessary in order to restrict both $\Delta_\sigma$ and $\Delta_\epsilon$. In some sense it's surprising this problem does not arise in three or four dimensions. Our postulated fusion rule, which is clearly reliant on Virasoro symmetry, is likely not more than a very rough approximation to the true spectrum of low-lying percolation operators in $D > 2$.  Even without finding a solution to  \eqref{constraint} in $6D$, there's still a signature of the free field result. In Figure \ref{6DP}, $\log(z)$ curves flatten as $\Delta_\sigma=2,\Delta_\epsilon=4$ is approached. The diminishing peaks can be viewed as a lesser violation of crossing symmetry, with the smallest such violation (peak) occurring when $\Delta_\sigma=2.002$ (red curve in Figure \ref{6DP}). A plot of $z$ at fixed $\Delta_\sigma=2.002$ exhibits a slight but well-defined dip at $\Delta_\epsilon=4.003$, as shown in Figure \ref{6DP_2}.

Unlike in even spatial dimensions, in $D=3$ and $D=5$ the fusion rule \eqref{perc_spectrum} is not adequate to distinguish both the spin and energy field scaling dimensions. In $3D$, for any given $\Delta_\sigma$ a vanishing singular value is present, but no clear minimal $z$ is found as a function of $\Delta_\sigma$ and $\Delta_\epsilon$. This may be due to the similarity in operator content and close proximity of percolation, SAW, and the Ising model, as all three theories have spin field scaling dimensions clustered near $\Delta_\sigma = 0.5$ in three dimensions. In $5D$ no vanishing singular values are present when $M>N$.  In both cases we can still bootstrap one of the scaling dimensions given the other is held fixed. Taking $\Delta_{\sigma,3D}=0.4765$ and $\Delta_{\sigma,5D}=1.4718$ \cite{Gracey}, $\Delta_{\epsilon,3D}=1.615$ and $\Delta_{\epsilon,5D}=3.416$ are obtained using $N=5, M=6$ and $N=M=7$, respectively. Compiled in Table \ref{ptable} are all of our bootstrapped scaling dimensions for percolation.

As a final note before moving on to the SAW, we mention the work of \cite{Flohr} which argues many of the relevant observables of $2D$ percolation can be obtained within a conformal field theory with $c=-24$.  Without restating their argument, they find all the weights in the Kac table shift by $-1$, implying
\begin{eqnarray*}
\Delta_\sigma &= 5/48 &\rightarrow -91/48 \\
\Delta_\epsilon &= 5/4 &\rightarrow -3/4.
\end{eqnarray*}
Bootstrapping with longitudinal derivatives and  \eqref{perc_spectrum} with scaling dimensions shifted accordingly, for fixed $\Delta_\sigma = -91/48$ we obtain a clear solution at $\Delta_\epsilon = -0.728$ with $N=M=6$. The general agreement with $\Delta_\epsilon = -3/4$ lends further evidence that our fusion rule is not just coincidentally successful.

\begin{table}
  \begin{center}
    \caption{Percolation scaling dimensions. Bold values are calculated with the bootstrap, and adjacent values in parenthesis are either exact results ($D=2,D=6$) or calculated by Pad\'e approximant at four loops ($D=3, D=4, D=5$) \cite{Gracey}. In odd spatial dimensions we're unable to determine both $\Delta_\sigma$ and $\Delta_\epsilon$, and instead bootstrap with the referenced value of $\Delta_\sigma$.}
   
    \label{ptable}
    \begin{tabular}{ c c c}
    \hline \hline
      $D$  & $\Delta_\sigma$ &$\Delta_\epsilon$  \\
      \hline 
      $2$  & $\mathbf{0.101} \, (5/48)$ &$\mathbf{1.235}\,(5/4)$ \\
      $3$ &-\quad $(0.4765)$  &$\mathbf{1.615}\, (1.8849)$   \\
      $4$ & $\mathbf{0.997} \, (0.9523)$ &$\mathbf{2.557}\, (2.5549)$ \\
      $5$ & -\quad $(1.4718)$ &$\mathbf{3.416}\, (3.2597)$ \\
     $6$ &  $\mathbf{2.002} \, (2)$ &$\mathbf{4.003} \,(4)$ \\
      
      \hline \hline
    \end{tabular}
  \end{center}
\end{table}

\section{IV. Self-Avoiding Walk}

The energy operator for the 2D SAW corresponds to the primary field $\Phi_{1,3}$  with $h_{1,3}=1/3$,  and it has a null state at level 3 rather than level 2 as in percolation. Therefore one difference in operator content which may distinguish the two $c=0$ theories is the inclusion of the $[\Delta_\epsilon+2,2]$ descendant.   Another is the inclusion of the lowest lying $O(N)$ symmetric tensor $[\Delta_T,2]$, whose dimension $\Delta_T \rightarrow \Delta_\epsilon$ as $N\rightarrow 0$ \cite{Shimada}. We find $T$ essential in applying the bootstrap to the SAW. The primary purpose of this operator is to input $O(N)$ symmetry. Secondarily it fulfills the role the identity operator did for percolation: it introduces an OPE coefficient independent of the energy sector, which can roughly account for the ignorance of logarithmic features. Retaining the identity operator in the presence of $T$ is therefore redundant; we find $2D$ scaling dimensions change by less than $5\%$ if the identity operator is also included. As in percolation
 $[D, 2]$ and other descendants of the identity discarded to avoid the $c=0$ catastrophe. For SAW in $2 \leq D \leq 4$ we thus create $\mathbf{F}$ with the operators
\begin{equation}
[\Delta_\sigma,0] \times [\Delta_\sigma,0] =[\Delta_\epsilon,0] +[\Delta_T,2] + [\Delta_\epsilon+2,2] + [\Delta_T+2,4] + [\Delta_\epsilon+4,4]+ [\Delta_\epsilon+6,6] + \dots
\label{SAW_spectrum}
\end{equation} and the $M$ lowest order longitudinal derivatives of $F_{\Delta_\sigma,\Delta,l}$ with $m\geq 3$.  Bootstrapping in $2D$ with the above spectrum we're unable to distinguish a solution with $\Delta_T,\Delta_\epsilon$, and $\Delta_\sigma$ all left arbitrary. The SAW is more difficult to isolate than percolation due to the collision of $\Delta_\epsilon$ and $\Delta_T$. Taking $N=M=6$, fixing both $\Delta_T=0.667$ and $\Delta_\sigma=5/48$ finds $\Delta_\epsilon=0.666$. With just a single scaling dimension fixed, the minimization procedure is not as reliable as in the percolation case, often getting caught in a local rather than a global minima. Fixing only $\Delta_T=0.667$ tentatively finds $\Delta_\epsilon = 0.666$, $\Delta_\sigma = 0.101$.

\begin{figure}
\begin{center}
\includegraphics[width=0.65\textwidth]{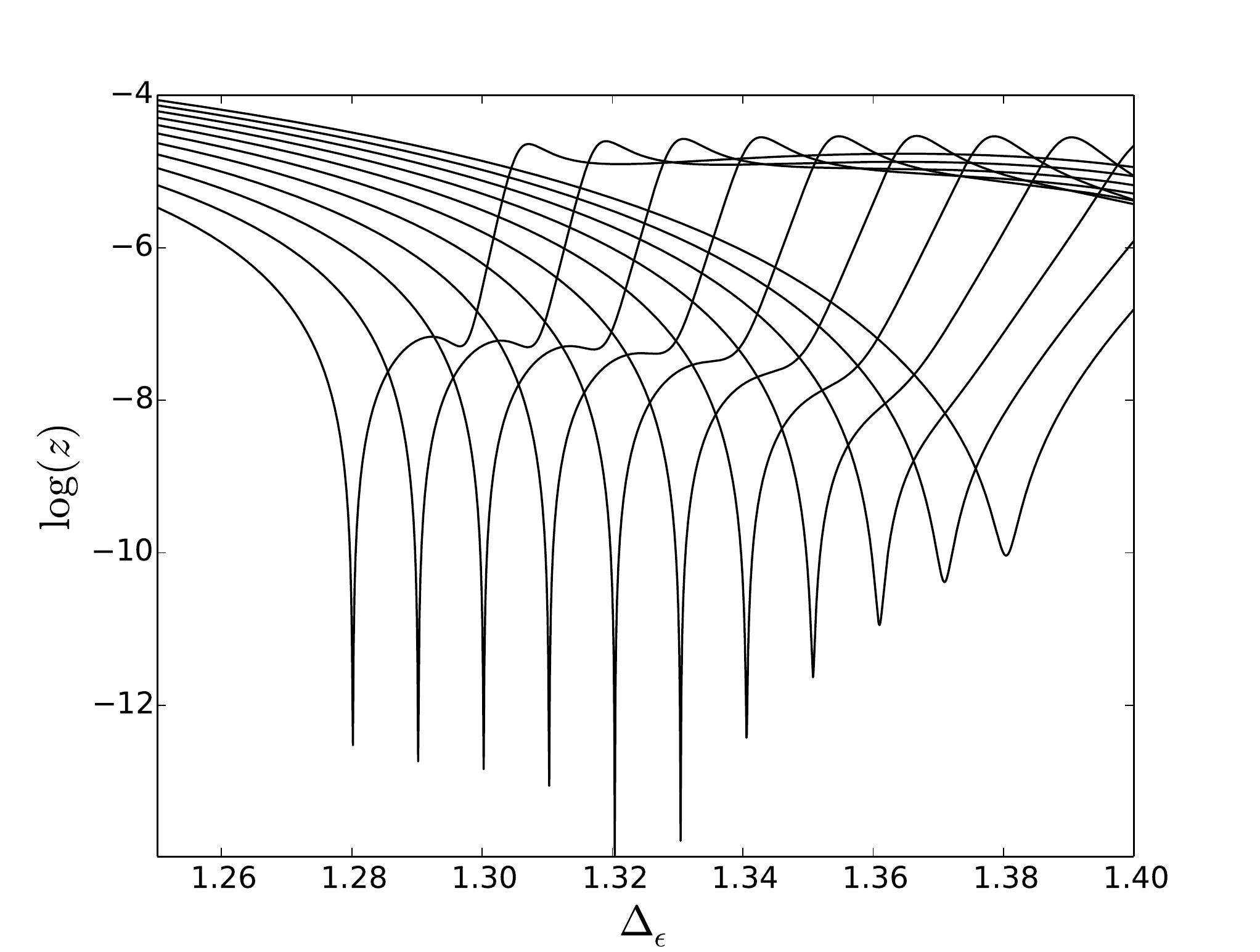}
\caption{{\bf 3D SAW.}   Logarithm  of the smallest singular value $z$ of $\mathbf{F}$ with $N=5, M=6$ as a function of $\Delta_\sigma$ and $\Delta_\epsilon$ for fixed $\Delta_\sigma=0.514$. Each curve corresponds to a distinct value of $\Delta_T$, linearly spaced from $\Delta_T = 1.28$ (left) to $\Delta_T=1.38$ (right). $\log(z)$ has a minimum at $\Delta_T = 1.326,  \Delta_\epsilon=1.326$. }
\label{3DSAW}
\end{center}
\end{figure}

\begin{figure}
\begin{center}
\includegraphics[width=0.65\textwidth]{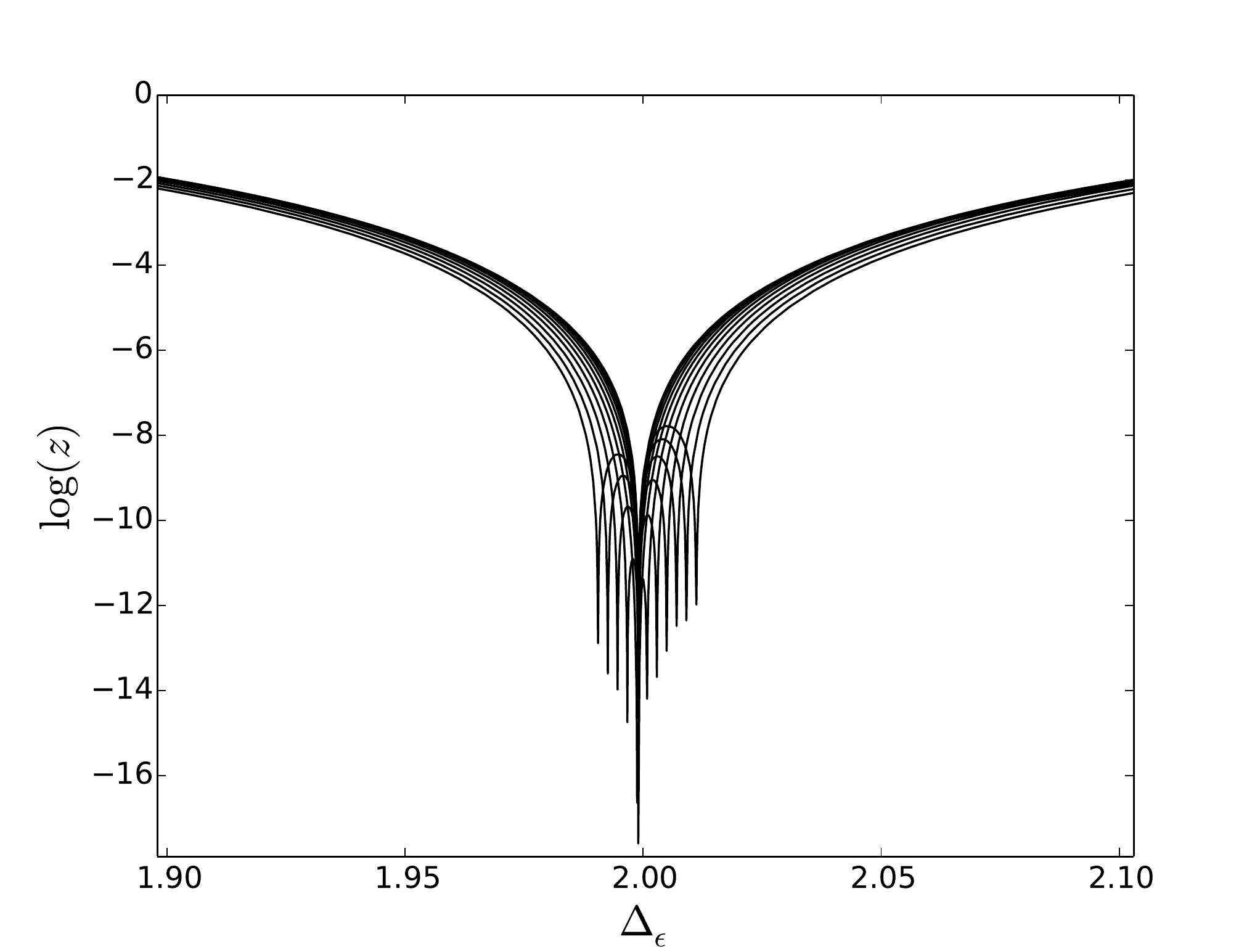}
\caption{{\bf 4D SAW.}   Logarithm of the smallest singular value $z$ of $\mathbf{F}$ with $N=6, M=8$ as a function of $\Delta_\sigma$ and $\Delta_\epsilon$ at $\Delta_T=1.999$. Each curve undergoes two dips in $\log(z)$, with one fixed at $\Delta_\epsilon=\Delta_T$ and the second shifting with $\Delta_\sigma$, which varies linearly from $ \Delta_\sigma=0.95$ (left) to $\Delta_\sigma=1.05$ (right). The two solutions coincide and achieve a minimal $\log(z)$ at $\Delta_T=1.999, \Delta_\epsilon = 1.999, \Delta_\sigma = 0.999$.}
\label{4DSAW_T}
\end{center}
\end{figure}

 In three and four dimensions the free theory obscures the SAW solution, due to the $[\Delta_\epsilon,0]$ and $[\Delta_T,2]$ operators. $\mathbf{F}$ always has a vanishing singular value as $\Delta_T \rightarrow \Delta_\epsilon$ because $G_{\Delta,2} \simeq G_{\Delta,0}$ near $\Delta=D-2$ where the scalar and spin two conformal blocks become degenerate. For the $3D$ SAW again there is difficulty in determining all three scaling dimensions using our proposed fusion rule \eqref{SAW_spectrum}. As we did for $3D$ percolation we fix $\Delta_\sigma=0.514$ \cite{Shimada} and bootstrap the remaining scaling dimensions using $N=5,M=6$, finding $\Delta_T=1.326$ and $\Delta_\epsilon=1.326$ (Fig. \ref{3DSAW}). In $4D$, with $\Delta_T, \Delta_\epsilon,\Delta_\sigma$ all arbitrary two solutions are present. One corresponds to $\Delta_\epsilon=\Delta_T$ and is independent of $\Delta_\sigma$. The second varies with $\Delta_\sigma$. Minimizing the smallest singular value of $\mathbf{F}$ as a function of $\Delta_T,\Delta_\epsilon$, and $\Delta_\sigma$ finds $\Delta_T=1.999, \Delta_\epsilon = 1.999, \Delta_\sigma = 0.999$ where the two solutions converge, as shown in Figure \ref{4DSAW_T}. This is expected, since the upper critical dimension for the self-avoiding walk is $D=4$. All bootstrapped SAW scaling dimensions are collected in Table \ref{stable}.

\begin{table}
  \begin{center}
    \caption{Polymer scaling dimensions. Bold values are calculated with the bootstrap, and adjacent values in parenthesis are either exact results ($D=2,D=4$), computed by $\epsilon$-expansion ($\Delta_\epsilon$ in $D=3$) \cite{Wilson}, or Borel summation ($\Delta_\sigma$ in $D=3$) \cite{Zinn}. }
   
    \label{stable}
    \begin{tabular}{ c c c c}
    \hline \hline
      $D$  & $\Delta_\sigma$ & $\Delta_T$ &$\Delta_\epsilon$  \\
      \hline 
      $2$  & $\mathbf{0.101} \, (5/48)$& - \, $(2/3)$ &$\mathbf{0.666}\,(2/3)$ \\
      $3$ &-\, $(0.514)$  &  $\mathbf{1.326}$\,$(1.336)$ &$\mathbf{1.326}$ \,$(1.336)$ \\
      $4$ & $\mathbf{0.999} \, (1)$ & $\mathbf{1.999} \, (2)$&$\mathbf{1.999}\, (2)$ \\

      \hline \hline
    \end{tabular}
  \end{center}
\end{table}

\vfill\eject

 \section{V. Summary}
The primary purpose of this work was to determine whether or not percolation and the self-avoiding walk could be distinguished with the conformal bootstrap. Though both theories share the same fusion algebra, central charge, and spin field scaling dimension in $D=2$, we've shown they can be isolated. Using a simplistic spectrum of operators based on Virasoro symmetry, and excluding descendants of the identity to indirectly specify $c=0$, the identity operator and a pair of spin $2$ operators -- a descendant of $\epsilon$ at level $2$ and an $O(N)$ symmetric tensor operator whose scaling dimension becomes degenerate with that of $\epsilon$ as $N\rightarrow 0$ -- can be used to discriminate between percolation and the SAW in any $D$.

For percolation in two and four spatial dimensions, our bootstrapped scaling dimensions agree relatively well with established results. In particular in $4D$ our determination of the correlation length critical exponent $\nu$, obtained with only $N=5$ operators, is within about $0.1\%$ of the value obtained by an involved four-loop calculation \cite{Gracey}.  For the upper critical dimension in  $6D$, while no rigorous bootstrapped solution is found we do see evidence of the anticipated free field solution. Bootstrapping percolation in odd $D$ is not as robust; to obtain a solution with our particular set of selection rules $\Delta_\sigma$ must be used as input. We point out that while this is the first treatment of percolation in $D>2$ with the conformal bootstrap, a similar implementation has been used to extract the structure constants of $2D$ percolation \cite{Picco}. Applying the bootstrap to the SAW,  for the upper critical dimension $4D$ we easily recover the expected scaling dimensions of the free theory. However, in $D=2$ and $D=3$ additional input is required to find solutions. Namely at least one of the three independent scaling dimensions appearing in the truncated spectrum must be held fixed. To conclude, while more accurate results are surely possible by using larger, more complicated spectrums, percolation and the self-avoiding walk are clearly distinguishable with the conformal bootstrap.

Encouraged by these results,  it would be interesting to use the conformal bootstrap to explore the space of $c=0$ theories in a systematic manner,   since many such theories are expected to have important physical applications.  In  particular very interesting problems in Anderson localization,   such as the elusive critical point for transitions in the  integer quantum Hall effect,   are expected to be described by a $c=0$ CFT in 2D \cite{Mirlin}.

\section{Acknowledgments}
We thank Tom Hartman, David Poland, and Gesualdo Delfino for encouraging discussions.

\vfill\eject

\section{A. Percolation Fusion Rule}
A potential criticism of our work is that to be in accordance with the exact fusion rule  $[ \sigma ] \times [\sigma] = [\epsilon]$, the identity operator's contribution should vanish at a solution if that solution is to truly represent percolation. In practice we instead find that the OPE coefficient of the identity, though minimized at a solution, is larger than that of the energy operator. In this appendix we posit that, while physically the identity operator should decouple, its inclusion is a) a numerical necessity in treating percolation with global conformal blocks in the Gliozzi bootstrap, and b) does not alter the bootstrapped scaling dimensions.

To show this, we'll consider $2D$ percolation. In $2D$ the 4-pt function can be written in terms of the Virasoro conformal blocks
\begin{equation}
\langle \mathcal{O}(\infty) \mathcal{O}(1) \mathcal{O}(z) \mathcal{O}(0) \rangle =  \sum_p a_p|\mathcal{F}(c,h,h_p,z)|^2.
\end{equation} Here $a_p$ are the OPE coefficients squared (note in general $a_p \neq p_{\Delta,l}$ \cite{Perlmutter}), $\mathcal{F}$ the Virasoro conformal blocks, and the sum runs over Virasoro primaries.  The utility of the Virasoro blocks for our purposes is twofold. First, each block contains all contributions to the four point function from a given conformal family, leading to simplification of the bootstrap equations for fusion rules containing just one Virasoro primary. Second, they're a function of $c$ and thus $c = 0$ can be implemented directly.
 
 To bootstrap with the Virasoro blocks, the analogues of the formulas provided in section II of the main text are required. These are provided in \cite{Esterlis}, for example, and restated here.
Crossing symmetry is respected if
\begin{equation}
\sum_p a_p^2\left[\mathcal{F}(c,h,h_p,z)\overline{\mathcal{F}}(c,h,\overline{h}_p,\overline{z})-\mathcal{F}(c,h,h_p,1-z)\overline{\mathcal{F}}(c,h,\overline{h}_p,1-\overline{z})\right]=0.
\end{equation} Expanding around $z=\overline{z}=1/2$ generates the homogeneous system
\begin{equation}
\sum_p a_p^2 \, g_{h,h_p}^{(m,n)} = 0
\label{vircrossing}
\end{equation} with
\begin{equation}
g_{h,h_p}^{(m,n)}  = \partial_z^m \partial_{\overline{z}}^n \left[\mathcal{F}(c,h,h_p,z)\overline{\mathcal{F}}(c,h,\overline{h}_p,\overline{z})-\mathcal{F}(c,h,h_p,1-z)\overline{\mathcal{F}}(c,h,\overline{h}_p,1-\overline{z})\right]|_{z=\overline{z}=1/2}.
\label{g}
\end{equation} Note $m+n$ must be odd or else $g_{h,h_p}^{(m,n)}$ is trivially zero. For the fusion rule $[\sigma] \times [\sigma] = [\epsilon]$ the homogeneous system becomes
\begin{equation}
\partial_z^m \mathcal{F}(c,h_\sigma,h_\epsilon,z)|_{z=1/2} = 0 \quad \text{or}\quad  \partial_z^n \mathcal{F}(c,h_\sigma,h_\epsilon,z)|_{z=1/2} = 0.
\label{simple_crossing}
\end{equation} As argued in \cite{Esterlis}, since $m+n$ is odd, either all even or all odd derivatives vanish at a solution to the crossing equation.

The argument above implies a simple way to determine whether or not it's even possible to use the Gliozzi bootstrap to find a solution with the correct OPE coefficients for 2D percolation: since either all odd derivatives or all even derivatives must vanish at a solution, if $\partial_z^1 \mathcal{F}(c,h_\sigma,h_\epsilon,z)|_{z=1/2} \neq 0$ and $\partial_z^2 \mathcal{F}(c,h_\sigma,h_\epsilon,z)|_{z=1/2} \neq 0$ as $c\rightarrow 0$ near $(h_\sigma = 5/96, h_\epsilon = 5/8$), then percolation can't be correctly found by the conformal bootstrap without treating the logarithmic CFT aspects more carefully.
 
\begin{figure}
  \centering
  \subfigure[\,$c=10^{-6}$]{\includegraphics[width=.49\textwidth]{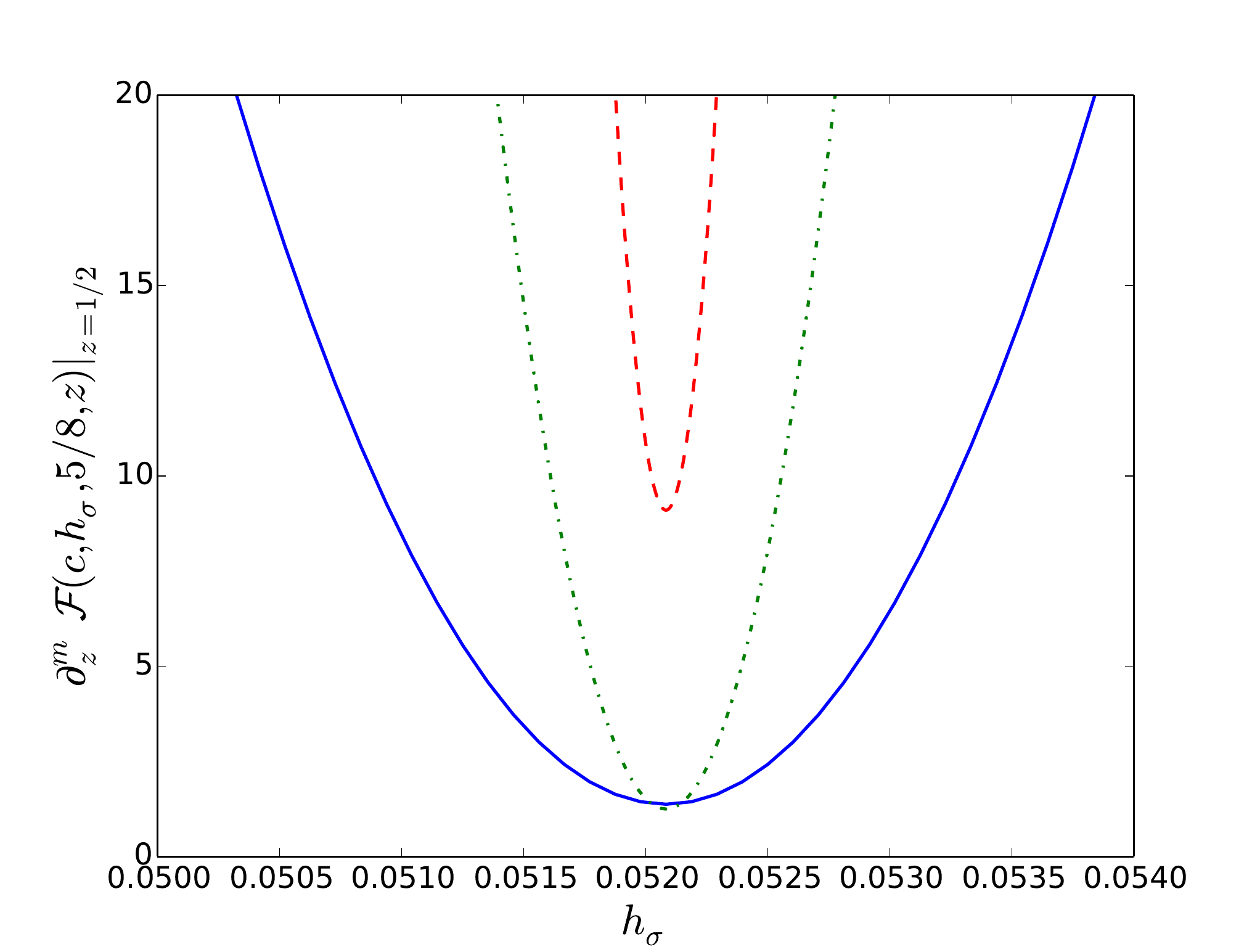}}\hfill
  \subfigure[\,$c=-10^{-6}$]{\includegraphics[width=.49\textwidth]{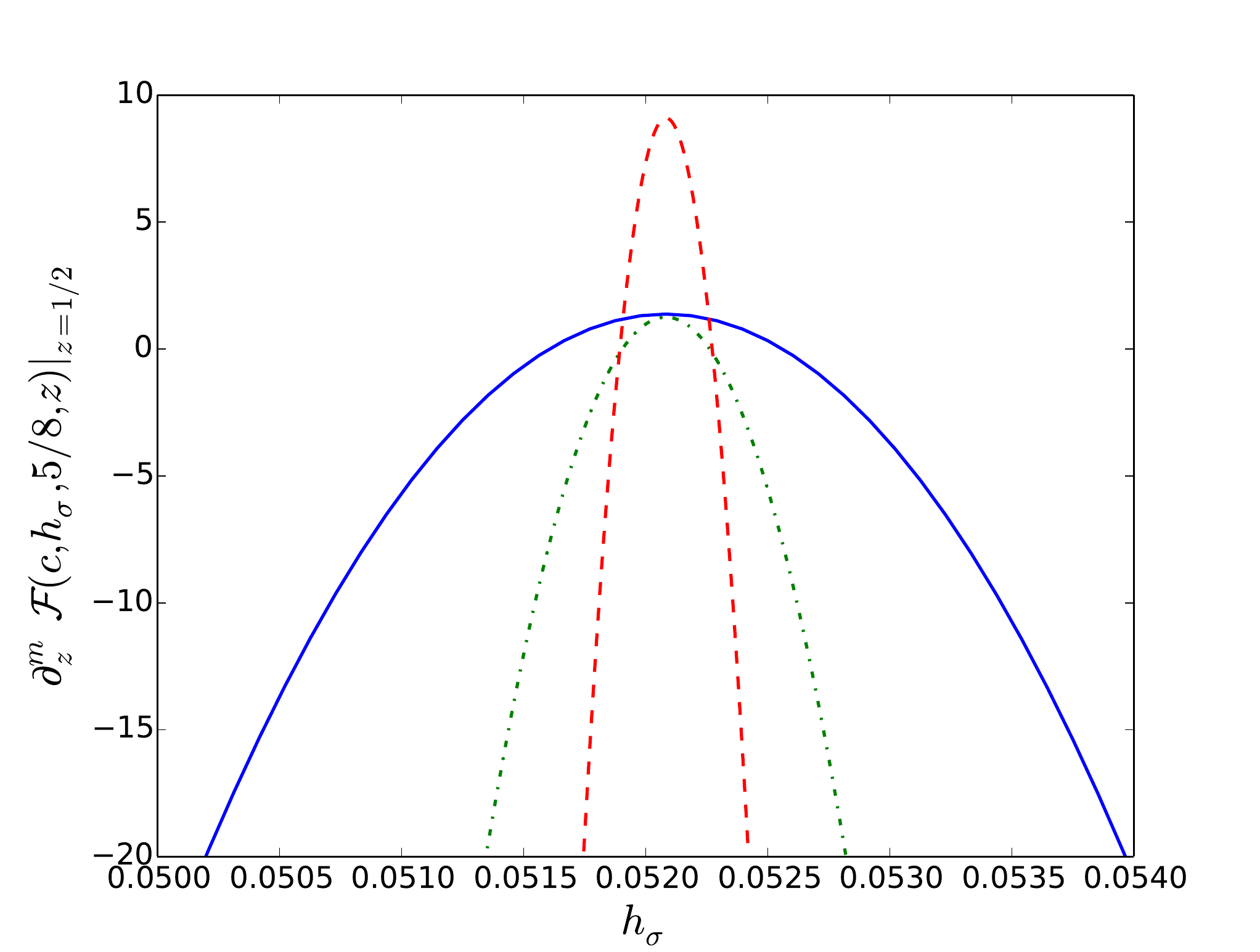}}
  \caption{$\partial_z^m \mathcal{F}(c,h_\sigma,5/8,z)|_{z=1/2}$ for $m=1,2,3$ (solid blue, green dash-dot, dashed red)}
  \label{app_perc_clims}%
\end{figure}

\begin{figure}
\begin{center}
\includegraphics[width=0.65\textwidth]{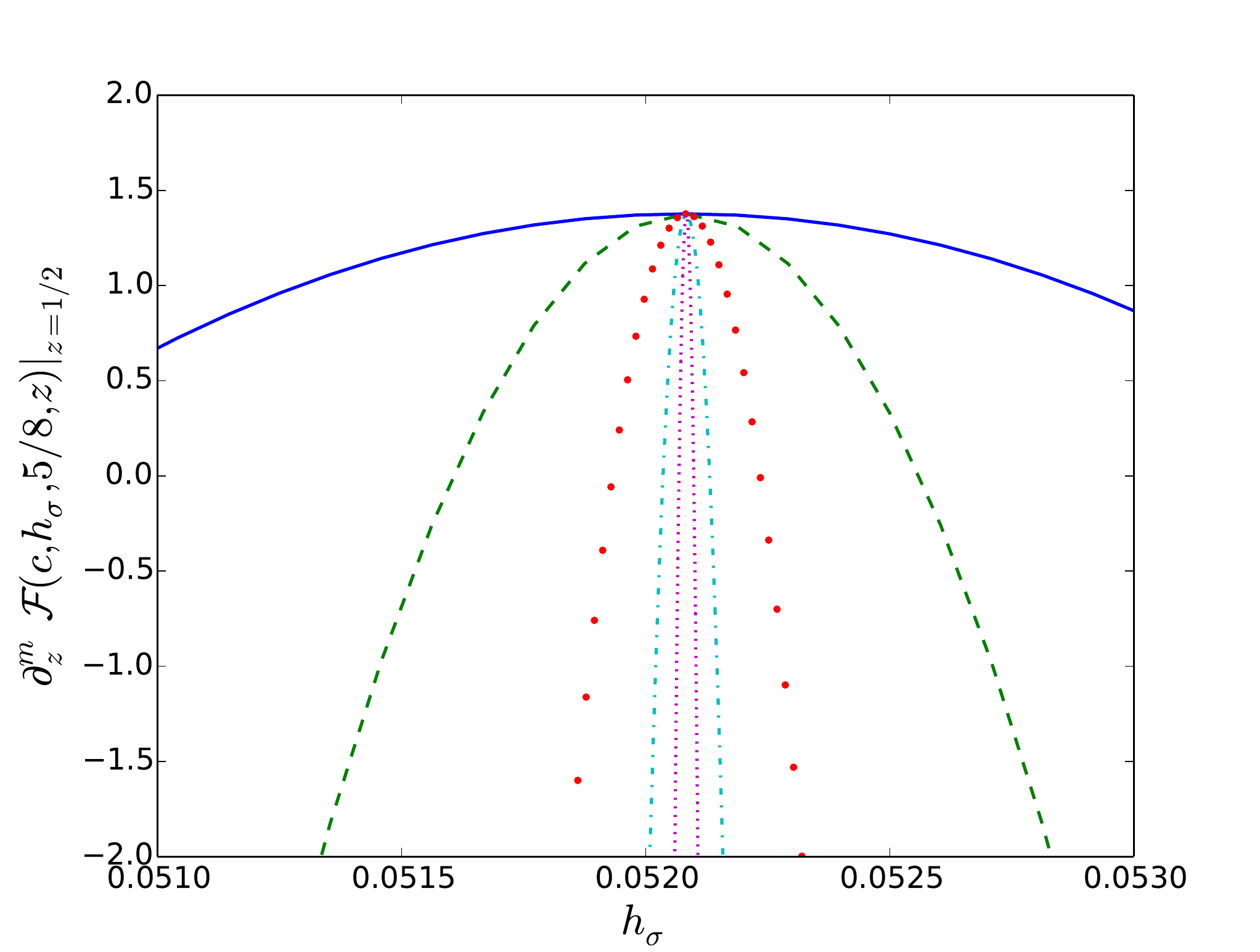}
\caption{$h_\sigma$ vs $\partial_z^1 \mathcal{F}(c,h,5/8,z)|_{z=1/2}$. The two solutions converge towards $h_\sigma = 5/96$ as $c\rightarrow 0$. $c$ values:$ -10^{-5}$ (solid blue),$-10^{-6}$ (dashed green),$-10^{-7}$(red circles),$-10^{-8}$ (cyan dash-dot),$-10^{-9}$ (magenta dots).}
\label{app_clim}
\end{center}
\end{figure}

The results (Fig \ref{app_perc_clims}) are unfortunately not so clear. With $h_\epsilon = 5/8$ fixed,  for $c>0$, no solution is found regardless of how close $c$ is to zero. For $c<0$, both even and odd derivatives vanish at two points equidistant from $h_\sigma = 5/96$. The two solutions converge as $c\rightarrow 0$, as shown in Fig. \ref{app_clim} for $\partial_z^1 \mathcal{F}(c,h_\sigma, 5/8,z)$. This structure is present only very near to $h_\epsilon=5/8$. This is expected; away from $q=1$ the fusion rule becomes $[\sigma] \times [\sigma] = [1]+[\epsilon]$.

The minima (maxima) of the $c>0$ ($c<0$) curves in Fig. \ref{app_perc_clims} all occur exactly at $h_\sigma = 5/96$, and clearly should correspond to $\partial_z^m \mathcal{F}(c,h_\sigma,h_\epsilon,z)|_{z=1/2}=0$ since percolation should be a solution. The shift above $0$ (which does not change as $|c| \rightarrow 0$) might represent the error in ignoring logarithmic terms in the OPE, which would be compounded in higher order derivatives. Including the identity operator in the fusion rule appears to correct the shift shown in Fig. \ref{app_perc_clims} at the cost of obtaining the correct OPE coefficients. In this case the sum in \eqref{vircrossing} contains $N=2$ terms: the $\epsilon$ block and the identity block.  The latter is given by the Virasoro vacuum block, truncated to include only the lowest order contribution
\begin{equation}
\mathcal{F}(c,h,0,z) = 1/z^{2h}.
\end{equation}

Since now two blocks are included in the fusion rule, the bootstrap must be performed with \eqref{g} rather than \eqref{simple_crossing}. With $m+n$ necessarily odd, we take $M=2$ derivatives and 
\begin{equation}
d_{23}=\begin{pmatrix}
g_{h,0}^{(2,1)} &  g_{h,5/8}^{(2,1)} \\
g_{h,,0}^{(3,0)} &  g_{h,5/8}^{(3,0)}
\end{pmatrix}
\label{d23}
\end{equation}with $c=-10^{-6}$ in order to make the closest possible comparison to the green and red curves of Fig. \ref{app_perc_clims}b. The smallest vanishing singular value of $d_{23}$ is found to occur at $h_\sigma=0.0519 \approx 5/96$.  Thus the two solutions equidistant from $h_\sigma = 5/96$  found with the exact fusion rule  are replaced with a single solution at the proper value solely by including the identity operator. The exact fusion rule is sufficient to find percolation only if $c\rightarrow 0^-$, which is unenforceable when bootstrapping with global blocks as in the main text. Keeping the identity operator in our fusion rule is essentially a numerical crutch; a method of correcting for using scalar rather than logarithmic conformal blocks. 

The drawback of retaining the identity operator in our fusion rule comes in the form of inaccurate OPE coefficients. For $a_\epsilon$ normalized to unity, inserting our solution ($h_\sigma = 0.0519, h_\epsilon = 5/8$) into the linear system associated with $d_{23}$ finds $a_\mathds{1} = 0.453$. As the magnitude of $c$ is further decreased this OPE coefficient grows, becoming larger than $a_\epsilon$. For example generating $d_{23}$ with $c=-10^{-7}$ instead leads to an approximate solution at $h_\sigma = 0.0521$ and $a_\mathds{1} \approx 4.3$. It's encouraging that even as $a_\mathds{1}$ increases  $h_\sigma$ remains relatively unperturbed. Including the identity operator here, and in the main text, does not drive the solution away from the percolation critical point. In this case its non-vanishing contribution, and more specifically $a_\mathds{1} > a_\epsilon$,  appears to be a signature of bootstrapping very close to $c=0$. This analysis suggests deviation from the known exact $h_\sigma, h_\epsilon$ values of $2D$ percolation has more to do with the truncation of the $\epsilon$ block than the presence of the identity operator. Also playing a role is the choice of derivative constraints used to construct the homogeneous system of equations in the bootstrap, which is the subject of Appendix B.

\section{B. Derivatives}

If a theory is easily truncable, which Taylor expansion terms are chosen to create $\mathbf{F}$ shouldn't strongly influence the outcome of the bootstrap. With the small number of operators kept in this work, a significant volatility in convergence is observed as the chosen set of derivatives is changed. This also arises in \cite{Hikami2} where for the $3D$ self-avoiding walk $\Delta_\epsilon = 1.325$ is found with just one of the four $3 \times 3$ minors considered.
 
\begin{figure}
  \centering
  \subfigure[\, $m \geq n$ derivative prescription finds a solution at $\Delta_\epsilon=0.667$, consistent with $2D$ SAW.]{\includegraphics[width=.49\textwidth]{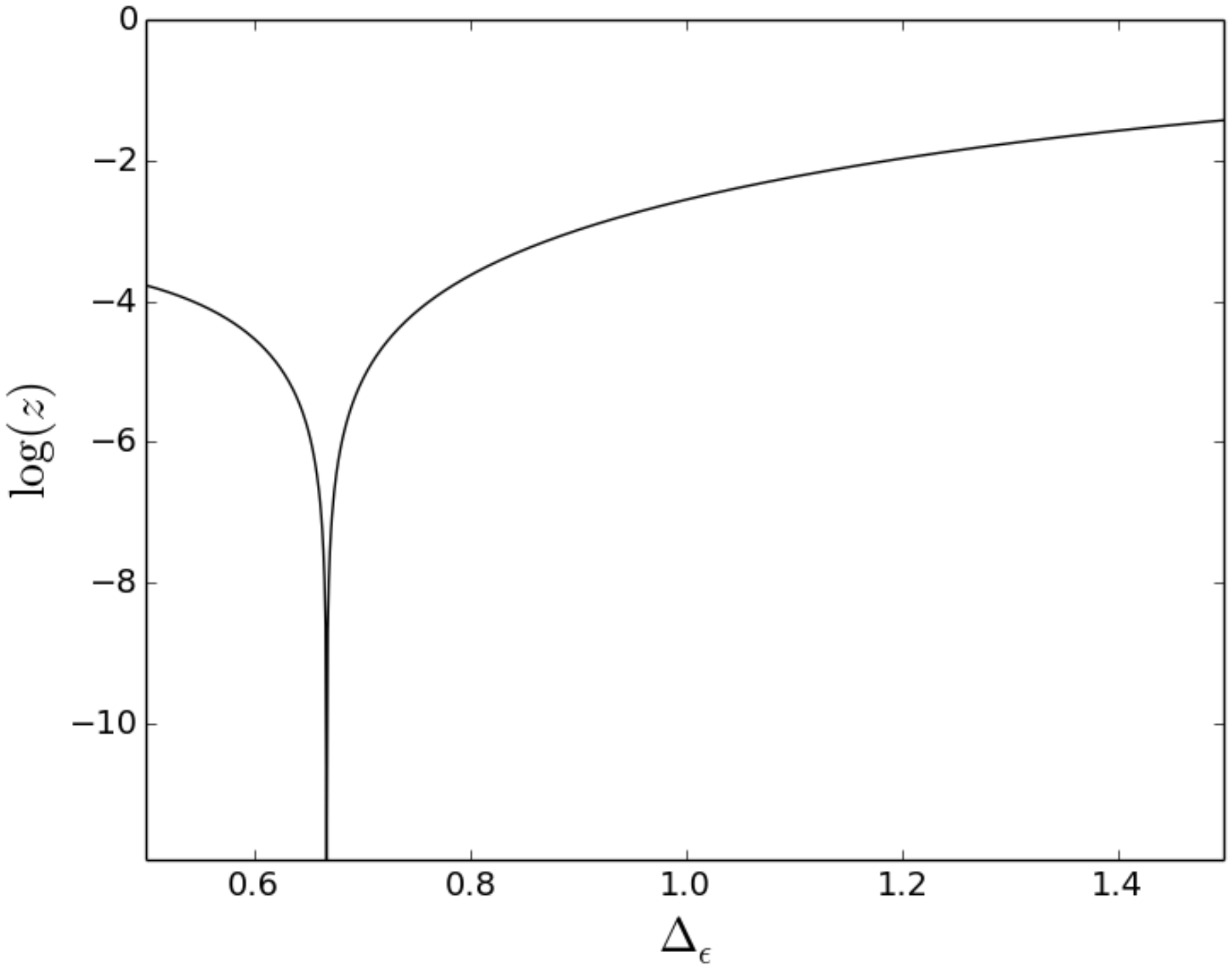}}\hfill
  \subfigure[\, $m \geq 3$ derivative prescription finds a solution at $\Delta_\epsilon=1.255$, consistent with $2D$ percolation.]{\includegraphics[width=.49\textwidth]{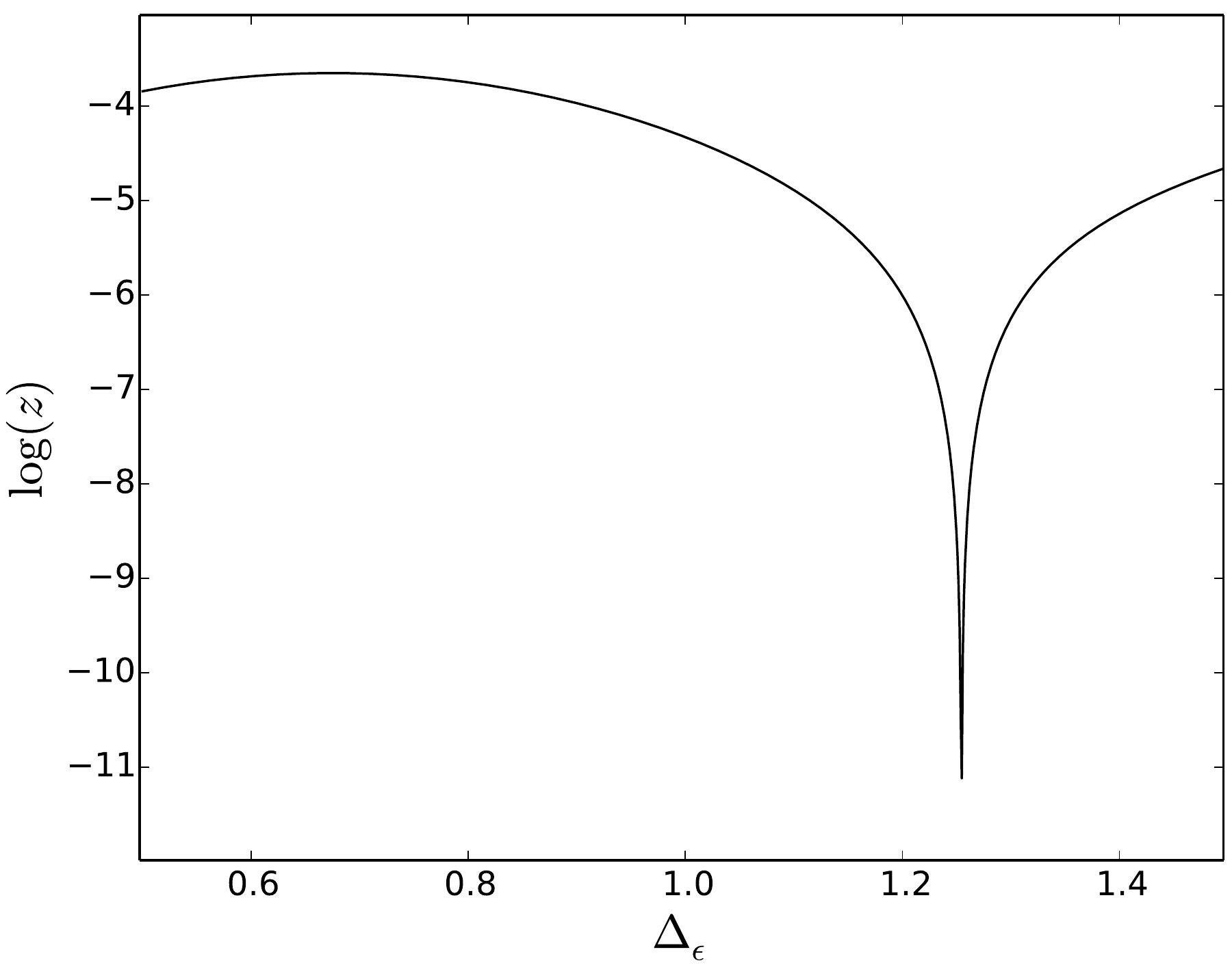}}
  \caption{Logarithm  of the smallest singular value $z$ of $\mathbf{F}$ for $N=M=6$ and fixed $\Delta_\sigma=5/48$.}
  \label{2Dfixed}
\end{figure}

 To illustrate we consider the spectrum
 \begin{equation}
 [\Delta_\sigma,0] \times [\Delta_\sigma,0] =  [0,0] + [\Delta_\epsilon,0] + [\Delta_\epsilon+4,4] + [\Delta_\epsilon+6,6] + [\Delta_\epsilon+8,8]+ \dots
 \label{appendix_spectrum}
 \end{equation} in $2D$. Aside from the identity these operators are all present in both the SAW and percolation. With this fusion rule and fixed $\Delta_\sigma=5/48$, we report in Table \ref{deriv_table} the bootstrapped value of $\Delta_\epsilon$, located by minimizing the smallest singular value of the  crossing matrix as a function of $\Delta_\epsilon$ for three different methods of choosing derivative constraints. For the natural choice $m\geq n$ (i.e. the $(m,n)$ sequence $(1,0), (1,1), (3,0),(3,1) \dots$) a solution which converges to the $2D$ self-avoiding walk $\Delta_\epsilon = 2/3$ is found. On the other hand employing only longitudinal derivatives and excluding $M=(1,0)$ (i.e. the sequence $(3,0),(5,0),(7,0)\dots$) finds a $\Delta_\epsilon$ consistent with percolation,  as shown in Figure \ref{2Dfixed}.
 
\begin{figure}
\begin{center}
\includegraphics[width=0.65\textwidth]{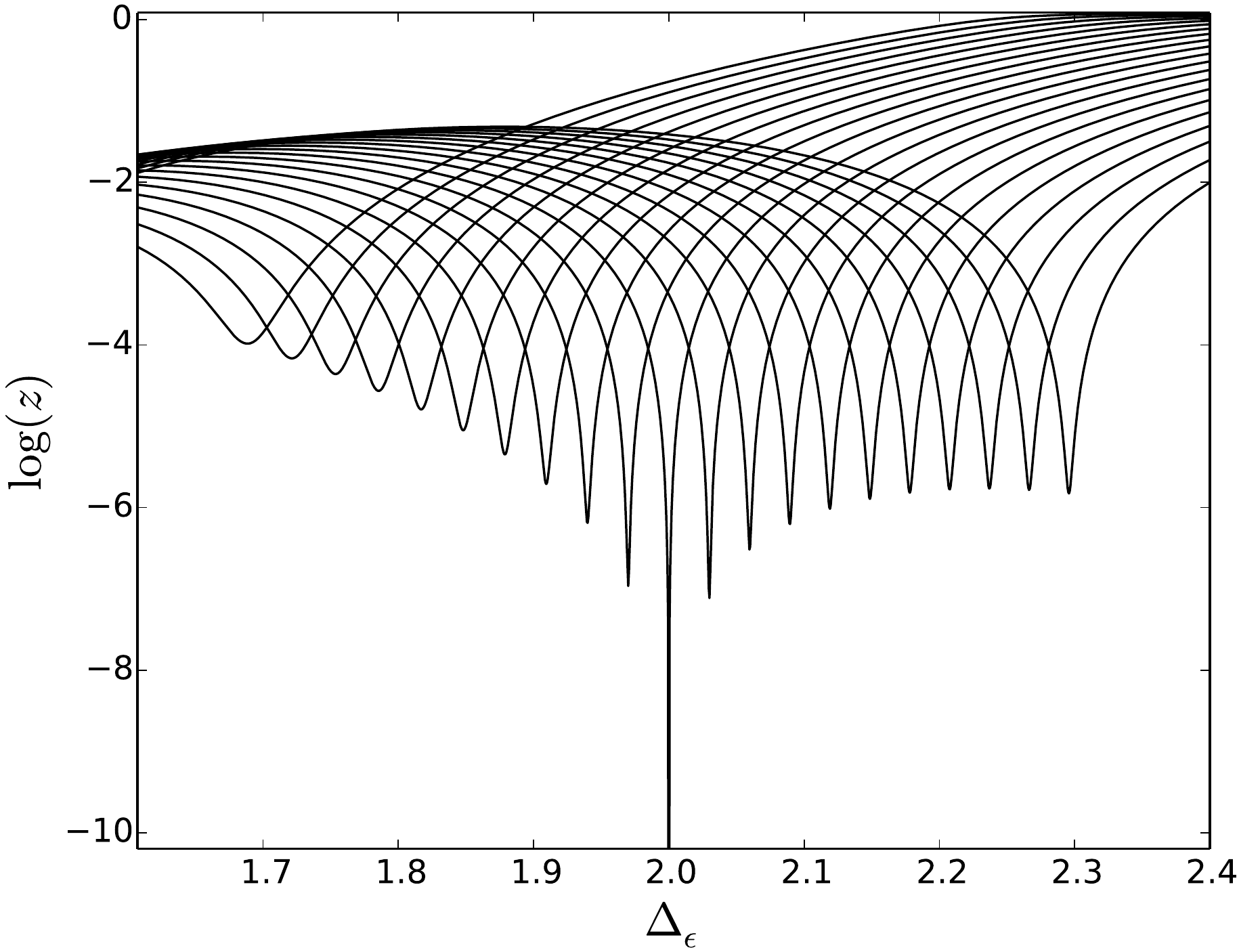}
\caption{ Logarithm of the smallest singular value $z$ of $\mathbf{F}$ with $N=6, M=7$ as a function of $\Delta_\sigma$ and $\Delta_\epsilon$, with the derivative prescription for 2D SAW. Each curve corresponds to a distinct value of $\Delta_\sigma$, linearly spaced from $\Delta_\sigma = 0.9$ (left) to $\Delta_\sigma =1.1$ (right). Minimizing $\log(z)$ finds the solution $\Delta_\sigma = 1.000, \Delta_\epsilon=2.000$ as anticipated for the $4D$ SAW.}
\label{4DSAW}
\end{center}
\end{figure}

Thus there is evidence polymers and percolation can be distinguished without appealing to the $O(N)$ symmetry of the self-avoiding walk as done in the main text, but instead by being selective with the Taylor expansion terms used to construct $\mathbf{F}$.  While this may appear to be just a trivial tuning of the system of equations to achieve a known result, using the same set of operators \eqref{appendix_spectrum} and the derivatives from column 2 (column 3) of Table \ref{deriv_table} also picks out percolation (SAW) in 4D, as shown in Figure \ref{4DP} (Figure \ref{4DSAW}). 

The decision to exclude transverse derivatives in the main text was initially made out of convenience; evaluating longitudinal derivatives of conformal blocks is less computationally intensive than evaluating their transverse counterparts. However, it's clear setting $n=0$ and using only longitudinal derivatives is more successful at bootstrapping $2D$ percolation. Presumably this variance in outcome, as shown in Table III, is evidence our spectrum of operators \eqref{perc_spectrum} is not comprehensive. With an exact, complete set of operators one would anticipate the results of the bootstrap being more robust. Indeed, when the fusion rule \eqref{appendix_spectrum} is expanded to include all descendants of the energy operator, which are inherently present in the $\epsilon$ Virasoro blocks making up $d_{23}$ in the previous appendix, utilizing the $(m,n)=(2,1)$ constraint is not a problem. Appendix A also sheds some light on why accuracy is improved if the $(m,n)=(1,0)$ term is avoided. In Fig. \ref{app_perc_clims}b the curves corresponding to $m=2$ and $m=3$ have solutions which exactly coincide while those of $m=1$ deviate further from $h_\sigma = 5/96$. This discrepancy is eliminated as $c\rightarrow 0^-$, but this isn't enforceable with global conformal blocks. In theory, implementing logarithmic conformal blocks \cite{Hogervorst2} along with increasing the number of retained operators should eliminate any need to worry about which derivative constraints are chosen.

\begin{table}
  \begin{center}
    \caption{Comparison of possible truncations of the crossing equation in two dimensions with fixed $\Delta_\sigma = 5/48$ and square matrices ($N=M$). In each successive row of the table, the lowest dimension operator from \eqref{appendix_spectrum} and lowest order derivative available is added, and the bootstrapped $\Delta_\epsilon$ is reported. Three possible methods of choosing derivatives are considered.}
   
    \label{deriv_table}
    \begin{tabular}{ c c c c }
    \hline \hline
     & $m\geq 1$ \quad & $m\geq 3$ \quad & $m\geq n$ \\
      $N$  &$(m,0)$  &$(m,0)$  &$(m,n)$  \\ 
      \hline 
      2  & 1.221 &1.321 & - \\
      3 & 1.216 &1.250 & 0.705   \\
      4 & 1.216 &1.260& 0.681 \\
     5 &  1.216 &1.255 & 0.672\\
      6 &  1.215 &1.255 & 0.667\\
      \hline \hline
    \end{tabular}
  \end{center}
\end{table}

\vfill\eject

\end{document}